\documentstyle[times,epsfig]{jaa}

\title[GMRT and VLA Observations of $l=24.8^{\circ}.8$, $b=0.^{\circ}1$]
{GMRT and VLA observations at 49cm and 20cm of the
HII region near $l=24^{\circ}.8$, $b=0.^{\circ}1$\\}
\author[N.G.Kantharia et al.]{N. G. Kantharia$^1$,
 W. M. Goss$^2$, D. Anish Roshi$^3$, 
\newauthor Niruj R. Mohan$^4$, Francois Viallefond$^5$ \\
$^1$National Centre for Radio Astrophysics,  TIFR,
Post Bag 3, Ganeshkhind, Pune-411007, India \\
$^2$National Radio Astronomy Observatory, P.O.Box 0, Socorro, NM 87801, USA \\
$^3$ Raman Research Institute, Sadashivnagar, Bangalore-560080, India \\
$^4$ Sterrewacht, Leiden, The Netherlands \\
$^5$Observatoire de Paris, Paris, France \\}

\begin{document}

\date{Received/Accepted 8 March 2007}
\maketitle

\begin{abstract} 

We report multifrequency radio continuum and hydrogen radio recombination 
line observations of HII regions near $l=24^{\circ}.8$ $b=0^{\circ}.1$ 
using the Giant Metrewave Radio Telescope (GMRT) 
at 1.28 GHz (n=172), 0.61 GHz (n=220) 
and the Very Large Array (VLA) at 1.42 GHz (n=166).
The region consists of a large number of resolved HII regions 
and a few compact HII regions as seen in our continuum maps, 
many of which have associated infrared (IR)
point sources.  The largest HII region at $l=24^{\circ}.83$ and $b=0^{\circ}.1$ 
is a few arcmins in size and has a shell-type morphology.  It is a massive
HII region enclosing $\sim 550~M_{\odot}$ with a linear size of 7 pc
and an rms electron density of $\sim 110$ cm$^{-3}$ at a kinematic
distance of 6 kpc.  The required ionization
can be provided by a single star of spectral type O5.5. 

We also report detection of hydrogen recombination lines 
from the HII region at $l=24.83^{\circ}$ and $b=0.1^{\circ}$ at all observed 
frequencies near $V_{lsr}=100$ kms$^{-1}$.  
We model the observed integrated line flux density as arising
in the diffuse HII region and find that the best fitting model has
an electron density comparable to that derived from the continuum.
We also report detection of hydrogen recombination lines 
from two other HII regions in the field. 
\end{abstract}


\section{Introduction}

Since the discovery of radio recombination lines (RRLs) 
(Dravskikh \& Dravskikh 1964 \nocite{dravskikh}; 
Sorochenko \& Borozich 1964 \nocite{sorochenko}; Palmer et al. 1967 \nocite{palmer})
these have been widely used
as diagnostics of ionized media in the Galaxy. Theoretical
studies of line formation have shown that the width of
RRLs depends sensitively on the principal quantum number
($\Delta \nu \propto n^{4.4}$; $\Delta \nu$ in Hz) 
and is also proportional to the electron density (Shaver 1975 \nocite{shaver2}). 
Thus low-frequency ($<$ a few GHz) RRLs from 
dense ionized regions ($>$ a few times 10$^2$ cm$^{-3}$) 
will be broadened, resulting in the reduction of the peak
line intensity. This reduction in peak intensity makes it difficult
to detect RRLs at low frequencies from high density gas. 
On the other hand, low frequency RRLs from low density ionized regions 
are not affected by this limitation. Hence these RRLs 
form an ideal probe to study lower density regions 
such as  photodissociation regions, diffuse, extended 
HII regions (e.g. Lockman 1989 \nocite{lockman}, 
Lockman et al. 1996 \nocite{lockman1}), envelopes of HII regions 
and the extended low density warm ionized
medium (ELDWIM) (Roshi \& Anantharamaiah 2001 \nocite{roshi1}, 
Anantharamaiah 1985 \nocite{anantha}, Heiles et al. 1996 \nocite{heiles}).
Typically such regions have densities $\le 100$ cm$^{-3}$
and emission measures $ \le 10^5$ pc~cm$^{-6}$.

In this paper, we present Giant Metrewave Radio Telescope (GMRT)
and Very Large Array (VLA) observations 
of the HII region complex near $l = 24^{\circ}.8$ and $b=0^{\circ}.1$ 
in RRLs and radio continuum at 
frequencies of 0.61, 1.28 and 1.42 GHz.  
This region consists of HII regions in various evolutionary stages, ranging from
protostellar stage showing bipolar outflows detected 
in CO (Furuya et al. 2002 \nocite{furuya}) to the diffuse, 
extended HII region like G24.83+0.10. The primary objective of these observations were
to image and determine the physical properties of the diffuse HII region from which
low frequency hydrogen and carbon RRLs near 327 MHz 
have been observed (Anantharamaiah 1985 \nocite{anantha}). The
details of the observations are given in Section~\ref{sec:obs}.  
The radio continuum morphology and physical properties of the diffuse HII regions
derived from the continuum are described in Section~\ref{sec:cont} . 
In Section~\ref{sec:rrl}, we present the RRL data. 
We end with a summary of the results.  A near kinematic distance of 6 kpc was
determined for G24.83+0.10 and is used in this paper. 
 
\section{Observations}
\label{sec:obs}

Table~\ref{tab0} gives the details of the observation. The 0.61 and 1.28 GHz
observations were carried out with the GMRT and the 1.42 GHz observations were
carried out with the VLA. 

\begin{table}
\caption{Observation details}
\begin{tabular}{lccc}
\hline
Parameter &  0.610 GHz & 1.28 GHz & 1.42 GHz \\ \hline \hline
Telescope &  GMRT      &  GMRT      & VLA (C, DnC) \\ 
Field of view & $\sim 44'$ & $\sim 25'$ & $\sim 40' $ \\
Phase centre &  &    & \\
~~~$\alpha_{2000}$ & $18^h36^m11^s$  & same & same\\
~~~$\delta_{2000}$ & $-07^\circ10'40''$  & same &same \\
Date of  & 17/1/2003 & 13/4/2002 & 5/10/2002, \\
Observation            &           & 14/4/2002 & 13/12/2002,\\
            &           &              &20/1/2003\\
Transitions      &  H$220\alpha$ & H$172\alpha$ & H$166\alpha$ \\
          &  C$220\alpha$ & C$172\alpha$ & C$166\alpha$ \\
Rest Freq  &    & &   \\
$~~~$ Hydrogen (GHz) &  0.613405  & 1.28117  & 1.42473   \\
$~~~$ Carbon (GHz) &  0.613711 & 1.28181  & 1.42544   \\
Phase cal    & 1822-096  &  1822-096          &     1831-126  \\
Flux cal & 3C286,3C48     &   3C286,3C48       &    1923+210   \\
Bandpass cal & 1822-096  & 3C286,3C48 &  3C48 \\
Bandwidth (km s$^{-1}$) & 480 & 480 &  330\\
Channel width (km s$^{-1}$) & 3.8 & 3.7 & 2.6 \\
Resolution & & &\\
~~~continuum& $14''\times7''$ & $8''\times6''$ & $35''\times25''$\\
~~~~~~~~PA     & $-24^{\circ}$ & $38^{\circ}$ & $40^{\circ}$\\
~~~line& $22''\times22''$ & $23''\times12''$ & $35''\times25''$\\
~~~~~~~~PA     & $-57^{\circ}$ & $42^{\circ}$ & $40^{\circ}$\\
RMS noise  & & &\\
~contm (mJy/beam) & 2  & 2 & 4\\
~line (mJy/beam) & 3  & 2 & 3\\
\hline
\end{tabular}
\label{tab0}
\end{table}

\subsection{GMRT observations: 0.61 and 1.28 GHz}

The GMRT (Swarup et al. 1991 \nocite{swarup} \& Ananthakrishnan \& Rao 2002 \nocite{ananth})
consists of 30 45-m diameter dishes spread over a 25 km region and
operating in five frequency bands.  
For our 1.28 GHz observations, we selected a 2 MHz bandwidth which includes both 
the H$172\alpha$ and the C$172\alpha$  lines. We followed an observing 
sequence of 20 min on-source followed by 6 min on 
the phase calibrator.  The bandpass calibrator was observed thrice during the run
for 30 min each. 
After editing corrupted data, we were left with 23 antennas on 13 April 2002 
and 17 antennas on 14 April 2002, which were used in the subsequent analysis.

At 0.61 GHz, we selected a bandwidth of 1 MHz 
which includes both the hydrogen ($H220\alpha$) 
and carbon ($C220\alpha$) lines.  
The observing procedure was similar to that used at 
1.28 GHz.  

The GMRT data
were converted to FITS format and imported into the NRAO AIPS package 
for further analysis. Single channel
data on the bandpass calibrators were first examined, 
edited and gain calibrated. Bandpass solutions were then generated on these
calibrators, which were applied to the data on all objects
before averaging the line free channels to generate the continuum database.
A first order spectral baseline was removed from the calibrated uv line data
using the task UVLIN.  The final continuum and line images were
generated using IMAGR.  The estimated flux density scale has errors $\le 20\%$.  

At the time of our observations, noise switching to measure
the system temperature was not available. We estimated the 
difference in system temperature between the primary
calibrator and the target source positions
using the 0.408 GHz continuum map (Haslam 1982 \nocite{haslam})
and assuming a spectral index for the galactic
background emission of $-2.7$. These correction factors were then
applied to the data.

\subsection{VLA observations: 1.42 GHz}

The VLA observations were conducted with a bandwidth of 1.5 MHz which
contained the observing frequencies for both the 
H$166\alpha$ and the C$166\alpha$ lines toward G24.83+0.10.
Observing details are listed in Table \ref{tab0}.  Data were obtained
in the C and DnC configurations. 
The observations on the target source were interspersed with 
$5$ min runs on the phase calibrator.  The bandpass calibrator was
observed at the beginning and end of the run for about $10$ min.  
Data from all the three short runs were combined to obtain the final dataset.
These data were online Hanning smoothed.
The VLA data were analysed using standard tasks in AIPS.
The continuum image was made by using the central
75\% of the total number of channels.  Gain and bandpass calibration
were then carried out and applied to continuum and line data.  The continuum
emission was removed from the line uv data using UVLSF in AIPS.
This data was then used to generate a cube using IMAGR. 
For further analysis, moment maps were also generated.

\section{The Continuum Emission}
\label{sec:cont}

\begin{figure*}
\includegraphics[width=3.0in]{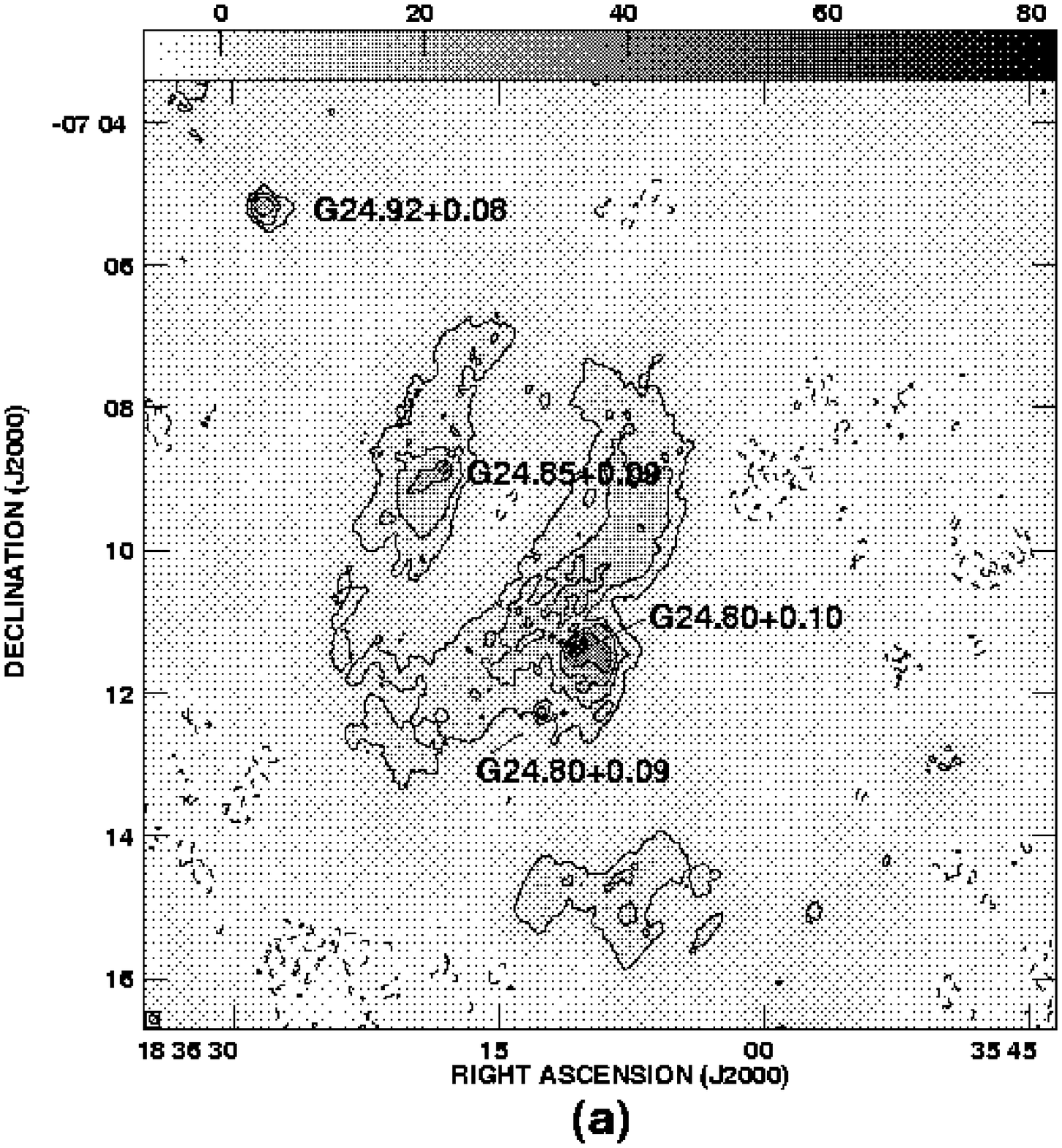}
\includegraphics[width=3.0in]{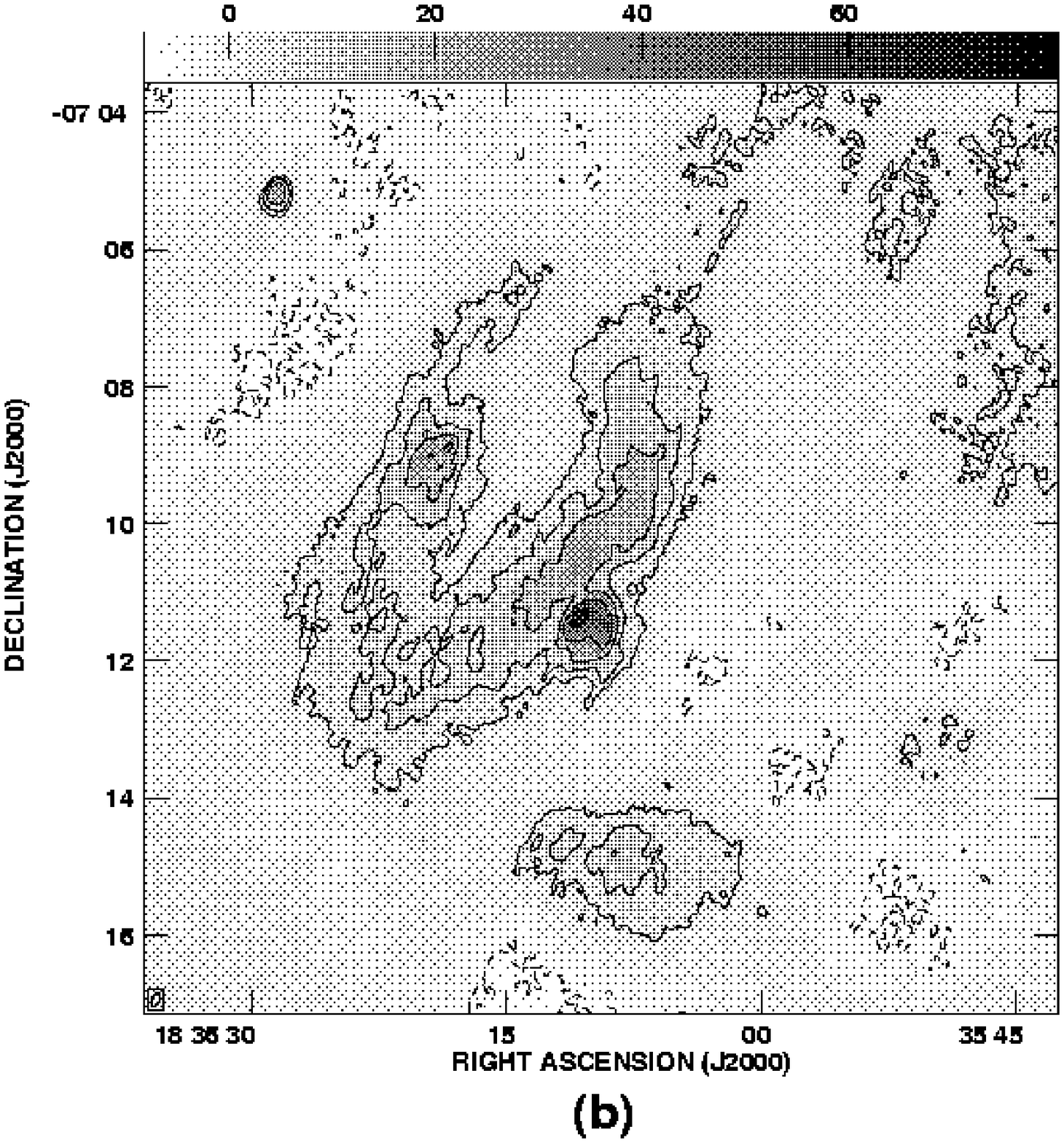}
\caption{(a) GMRT 1.28 GHz continuum image of the 
region near $l=24^{\circ}.8$, $b=0^{\circ}.1$ 
with an angular resolution of $8.4'' \times 6.3''$, 
PA $=38^{\circ}$.  The RMS noise in the image is 1.5 mJy/beam.  
The grey scale ranges from $-$7 mJy/beam to 82 mJy/beam. 
The contour levels are -7.5, -3.75, 3.75, 7.5, 15, 30, 60, 120, 240, 480 mJy/beam. 
The compact sources detected in the field are labelled. 
(b) GMRT 0.61 GHz continuum image of the region near $l=24^{\circ}.8$, 
$b=0^{\circ}.1$.  The angular
resolution of the image is $13.8''\times 6.7''$, PA $= -24^{\circ}$ and the RMS noise
is 1.6 mJy/beam.  The grey scale ranges from $-$8 mJy/beam to
80 mJy/beam.  The contour levels are -8, -4, 4, 8, 16, 32, 64, 128, 256, 512 mJy/beam.  }
\label{fig1}
\end{figure*}

\begin{figure}
\resizebox{\hsize}{!}{\includegraphics{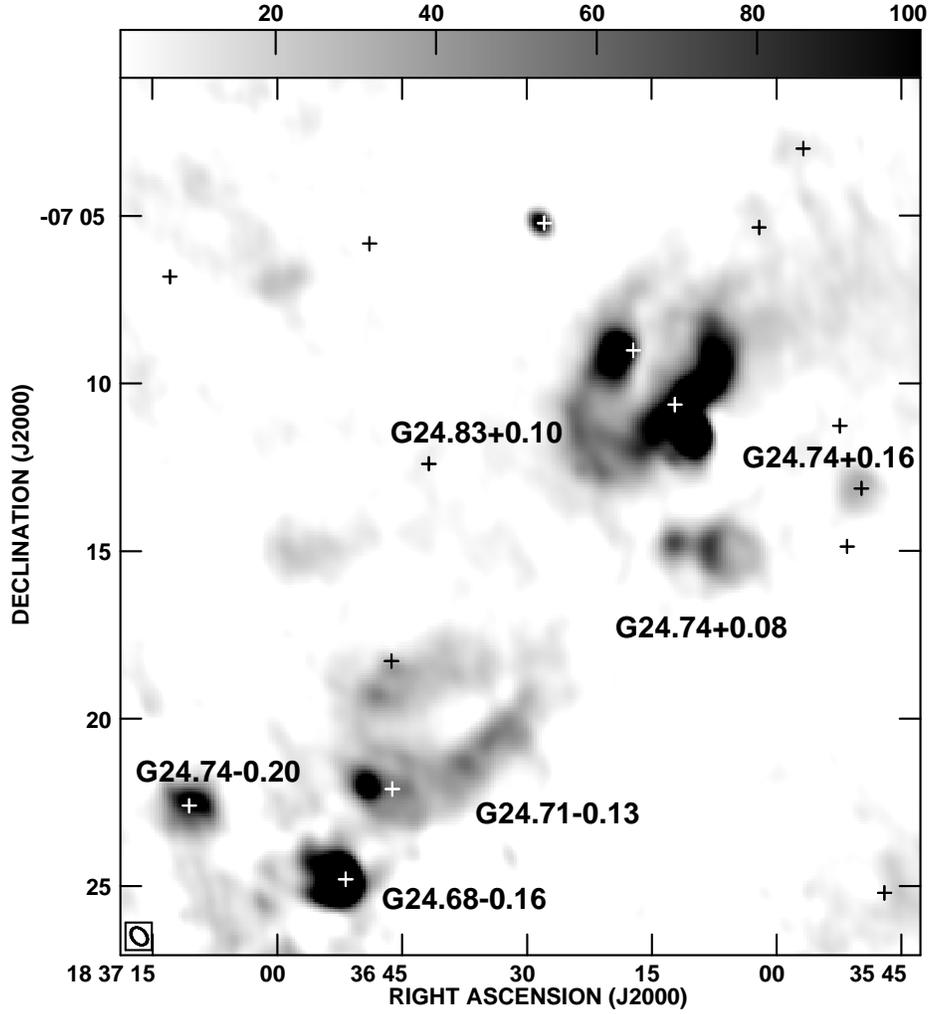}}
\caption{(a) VLA radio continuum image at 1.42 GHz of the field near
$l=24^{\circ}.8$, $b=0^{\circ}.1$ .  The image has been smoothed
to a resolution of  $35'' \times 25''$,  PA $= 40^{\circ}$.
The grey scale is plotted from 1 mJy/beam to 100 mJy/beam. 
The image has been corrected for primary beam attenuation.
The diffuse HII regions (angular size $\ge 1'$) in the field are 
labelled.  Crosses mark positions of IRAS point sources in the region obtained from the 
IRAS PSC.  }
\label{fig2}
\end{figure}

\addtocounter{figure}{-1}
\begin{figure}
\resizebox{\hsize}{!}{\includegraphics{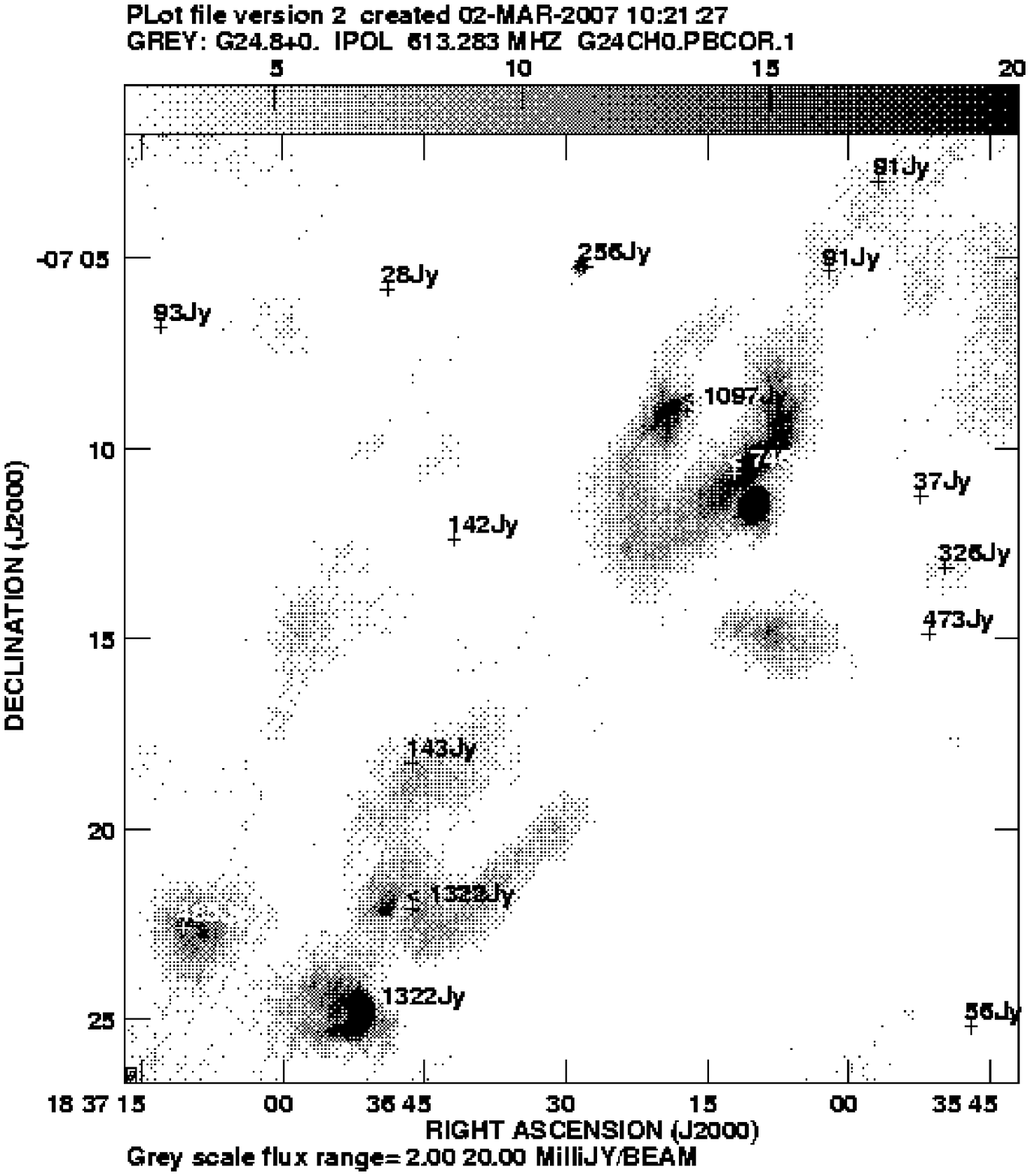}}
\caption{(b) GMRT image of the same field as in (a) at 0.61 GHz. 
The image has an angular resolution of
$14'' \times 7''$,  PA $= -24^{\circ}$.
The grey scale is plotted from 2 mJy/beam to 20 mJy/beam. 
The image has been corrected for primary beam attenuation.  The crosses mark
the positions of sources from the IRAS PSC.  The strength of these
sources at $60 \mu$m are marked. }
\label{fig2}
\end{figure}

\begin{table}
\caption{Continuum flux densities and derived spectral indices using
0.61 and 1.42 GHz data of the diffuse HII regions marked in Fig.~\ref{fig2}.  Since some of
the regions lie outside the 1.28 GHz primary beam, no flux density at
this frequency is available.   } 
\begin{tabular}{lcccc}
\hline
Source &  $\nu$ & Flux density$^1$   & $\alpha^{0.61GHz}_{1.42GHz}$ & IRAS Pt src \\
       &  GHz     & Jy   &    &  \\
\hline
 & & &  &  \\
G24.83+0.10  &   0.61  &  $8.4$  &  $-0.1^{-0.2}_{+0.3}$  & IRAS 18335-0713A \\
           &  1.28  &  $5.3 $  &    & IRAS 18335-0711 \\
           &  1.42  &  $7.7 $  &  & \\
           &  $4.87^2$  &  $7.1$  &   &   \\
&  & & & \\
G24.74+0.08      &  0.61  &  $1.29 $  & $-0.6^{-0.2}_{+0.3}$ &  -   \\
           &  1.28 &  $0.65 $   &  &  \\
           &  1.42 &  $0.80 $  &  &  \\
 & & & & \\
G24.74+0.16 & 0.61   &  $0.21$   &  $-0.4^{-0.2}_{+0.3}$  & IRAS 18331-0715   \\
           & 1.42  &  $0.15 $  &   &  \\
 &&  & & \\
G24.68$-$0.16 & 0.61   &  $2.6  $   &  $-0.15^{-0.2}_{+0.25}$ & IRAS 18341-0727    \\
           & 1.42  &  $2.3 $  &   &  \\
 &&  & &\\
G24.71$-$0.13 & 0.61 & 3.9  &  $-0.3^{-0.2}_{+0.25}$ & IRAS 18340-0724   \\
 & 1.42 & 3.0  & & \\
 &&  & & \\
G24.74$-$0.2 & 0.61 & 1.4   & $-0.85^{-0.15}_{+0.2}$  & IRAS 18344-0725\\
 & 1.42 & 0.68  && \\
 &&  & & \\
\hline
\end{tabular}

$^1$ The error in flux density at 0.61 GHz and 1.28 GHz is $\le 20\%$.  The
error bars on the spectral index are estimated using a conservative
$20\%$ error on the 0.61 GHz data points.  \\
$^2$ Data from Downes et al. (1980). \\
\label{tab1}
\end{table}

\begin{table}
\caption{The physical parameters derived from the measured continuum flux density
of the source at 1.42 GHz are listed below.  We assumed an electron temperature
of $7000$ K for the HII region and a spherical geometry.
The exciting star is tabulated from Panagia (1973). All
the regions are taken to be at a kinematic distance of 6 kpc.}
\begin{tabular}{lcccccc}
\hline
{\bf Source} &  ${\bf <N_{e}>}$ & {\bf d} & ${\bf <EM>}$ & {\bf Mass} & {$\bf N_{Lyc}$}
  &  Exc \\
 &  cm$^{-3}$ & pc & $10^4$ & $M_{\odot}$ & $10^{48}$  & star \\
&  & & pccm$^{-6}$  & & sec$^{-1}$  & \\
\hline
G24.83+0.10  & 110 & 7 & $9.0$ & 550 & $23$ & O5.5 \\
G24.74+0.08  & 90 & 4 & $3.5 $ & 70 & $2 $ & O8  \\
G24.74+0.16  & 75 & 3 & $1.5 $ & 15 & $0.4$ & B0 \\
G24.68-0.16  & 160 & 4 & $9.5$ & 120 & $7$ & O6.5 \\
G24.71-0.13 & 70 & 7 & 3.5 & 340 & 9  & O6 \\
G24.74-0.20 & 160  & 3 & 6.5 & 35 & 2 & O8  \\
\hline

\end{tabular}
\label{tab3}
\end{table}

Fig.~\ref{fig1} and \ref{fig2} show the continuum images.   
The 1.28 (Fig.~\ref{fig1}a) and 0.61 (Fig.~\ref{fig1}b, Fig. ~\ref{fig2}b) GHz continuum images 
were obtained after applying a UV taper of 30 k$\lambda$. 
At 1.28 GHz, the largest angular scale to which the GMRT is sensitive is  $\sim 7'$,
thus the images are likely missing flux density for the large scale components. 
We, therefore, use the 0.61 and 1.42 GHz data to derive the spectral index listed
in Table \ref{tab1}.
The compact sources are marked in the 1.28 GHz image (see Fig~\ref{fig1}a).
Resolved sources within the field of view of the observations are marked 
in the 1.42 GHz image shown in Fig.~\ref{fig2}a.  Table~\ref{tab1}
lists the estimated flux densities of these diffuse sources after correcting
for the primary beam attenuation.  IRAS point sources are marked by crosses
in Fig.~\ref{fig2}a. 
 
Infrared sources are associated with most of the diffuse regions 
shown in the Fig \ref{fig2}a.  
To ascertain that these are not chance superpositions, we examined the
60$\mu$m flux density of these sources as given in the IRAS PSC and these
are plotted in Fig.~\ref{fig2}.  Codella et al. (1994 \nocite{codella}) 
have shown that IRAS PSC sources which have $S_{60\mu}m> 100 $Jy
have more than $80\%$ probability of being associated with HII regions
and that for sources with $S_{60\mu}$m$ > 300 $Jy, chance superpositions are expected
to be very rare.  The  $S_{60\mu}$m for all the sources are plotted in
Fig.~\ref{fig2}b.  As seen here, almost all the sources lying on the diffuse
sources have a flux density $> 100$ Jy at $60 \mu$m and most of them have flux densities
$ > 300$ Jy totally ruling out chance superpositions.  The 
IRAS sources seen in rest of the field have lower flux densities at
$60 \mu$m and hence are unlikely to be associated with HII regions.  

In rest of this section, we discuss
the continuum emission from these diffuse sources alongwith the associated IRAS
point sources.  The radio continuum peaks in the diffuse regions are likely
indicating the positions of the exciting stars.

{\bf G24.83+0.10 :} The brightest diffuse source in the field of view
is G24.83+0.10. The integrated flux density of this source at 1.42 GHz
is 7.7 Jy. Higher resolution images of this source are shown 
in Fig~\ref{fig1}. 
The diffuse source has a shell-like morphology with multiple radio peaks. 
The size of the shell is about 7 pc.
The spectral index of this HII region is consistent with optically thin
free free emission.  Two IRAS point sources are coincident
with this object as shown in Fig.~\ref{fig2}.  
The IR emission arises from the thermal dust
located within the OB associations heated by the stars (Conti \& Crowther 
2004 \nocite{conti}).
At higher angular resolutions and frequencies, several compact/ultracompact HII regions
are detected in G24.80+0.09.  The ultracompact HII regions in G24.8+0.09 have been 
extensively studied
at higher frequencies with high angular resolution (for example,
Furuya et al. 2002 \nocite{furuya}).  These HII regions are found to be in 
an earlier stage of evolution (Furuya et al. 2002 \nocite{furuya}).
The UV photons from the stars exciting the UC HII regions might be absorbed by dust
resulting in photon-bound UC HII regions.
The other peak in the W arc, ie G24.80+0.10 appears to be a
cometary type HII region with a tail extending to the N
along the diffuse arc.  
The IRAS point source IRAS 18335-0713A is located in the W arc of G24.83+0.10, 
close to G24.80+0.10 and has a flux density of 746 Jy at 60 $\mu$m. 
G24.80+0.10 (which appears as a single radio peak in the VLA 20 cm
image (see Fig \ref{fig2})) is resolved into a double peaked
structure in the higher resolution GMRT images at 20 cm and 49 cm.
Another compact HII region G24.85+0.09 (see Fig~\ref{fig1}a) is located
in the E arc of this diffuse HII region.  This region also
seems to display a cometary structure. 
The IRAS point source IRAS 18335-0711 with a flux density of 
1097 Jy at 60$\mu$m is associated with this region.

We have used the VLA data at 1.42 GHz to estimate the average physical
parameters of the diffuse HII region. 
The RMS electron density, emission measure,
mass of the ionized gas and the number of Lyman continuum photons
required to maintain ionization are listed in Table \ref{tab3}
(using Mezger and Henderson 1967 \nocite{mezger}). The exciting star 
types, assuming a single ionizing star are listed in the last 
column of Table \ref{tab3} (Panagia 1973 \nocite{panagia}). 
G24.83+0.10 is the most massive HII region. 
In the next section, we discuss the
observed recombination line emission from this region.

{\bf G24.74+0.08 :} This diffuse region located due S of G24.83+0.10 (see Fig~\ref{fig2})
does not have an associated IRAS point source.
However, strong emission in the mid-infrared band E (MSX data) is observed, 
indicating its thermal nature and the predominance of hot dust.
Moreover, MIR band A (MSX data), dominated by emission from polycylic
aromatic hydrocarbons (PAH), is also detected from this region.  
The physical properties of this HII region estimated from its
continuum strength are listed in Table \ref{tab3}.

{\bf G24.71-0.13 :} This is another shell-like HII region.
The size of the shell is
about 7 pc  similar to that of G24.83+0.10.
G24.71-0.13 (see Fig~\ref{fig2}) has a radio continuum peak in the middle of the two arcs
forming the shell with diffuse emission along the rest of the shell.
The IRAS point source IRAS 18340-0724 of strength 1322 Jy at 60$\mu$m 
is located close to the radio peak and is associated with the HII region.
The properties of G24.71-0.13 derived
from the continuum observations are listed in Table \ref{tab3}. 

{\bf Other continuum sources :}The HII regions 
G24.74+0.16, G24.74-0.2 and G24.68-0.16 are relatively compact with an angular extent
less than $\sim$ 2'. All three have associated IRAS point sources.
G24.68-0.16 is the brightest among these
four sources (2.3 Jy at 1.42 GHz).  IRAS 18341-0727 with a flux density
of 1322 Jy at $60 \mu$m associated with it.  The emission is consistent with a free-free
spectrum.  This source appears to have a shell-like
morphology with the shell opening to the E.  G24.74+0.16 located to 
the W of G24.83+0.10 is the weakest continuum source in the field.  The IRAS point source
IRAS 18331-0715 of strength 326 Jy at $60 \mu$m
is associated with this HII region.  The derived physical properties of 
these HII regions are also listed in Table \ref{tab3}.
An unresolved HII region, G24.92+0.08 (see Fig \ref{fig1}a) is located to 
the NE of G24.83+0.10. This HII region has an associated
IRAS point source IRAS 18337-0707 with a strength of 256 Jy at $60 \mu$m. 

\section{The Recombination Line emission}
\label{sec:rrl}

\begin{figure*}
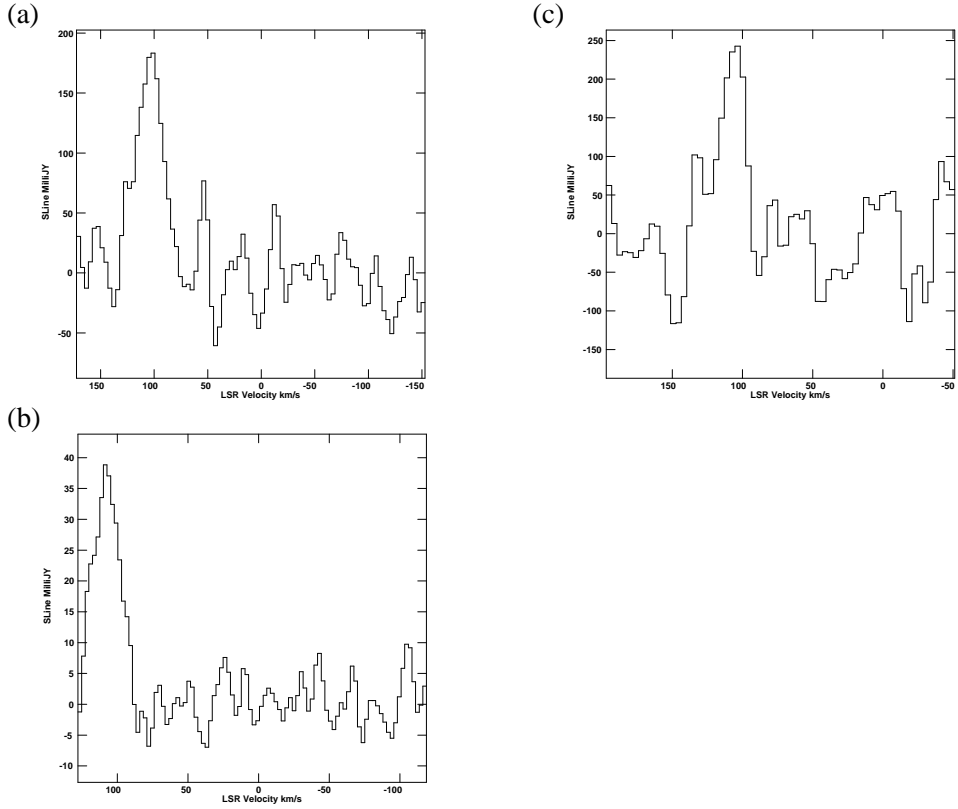

(a)\includegraphics[width=2in,angle=-90]{fig3a.ps}
(c)\includegraphics[width=2in,angle=-90]{fig3b.ps}
(b)\includegraphics[width=2in,angle=-90]{fig3c.ps}
\caption{{\bf (a)} The GMRT $172\alpha$ spectrum integrated over G24.83+0.10. 
The spectrum
has been Hanning smoothed to a velocity resolution of 7.2 kms$^{-1}$.
{\bf (b)} The GMRT $220\alpha$ spectrum integrated over the continuum source G24.83+0.10.
The spectrum has been Hanning smoothed
to a velocity resolution of 7.6 kms$^{-1}$.  
{\bf (c)} The VLA $166\alpha$ spectrum integrated over a small region in 
G24.83+0.10 near G24.80+0.10 located in the W arc. 
The spectrum has
been Hanning smoothed to a spectral resolution of 5.2 kms$^{-1}$.  
The feature near 100 kms$^{-1}$ in all the three figures are the H RRLs. } 
\label{fig3}
\end{figure*}

Hydrogen RRLs from G24.83+0.10 were detected
at all three observed frequencies.  Fig \ref{fig3} shows the spectra
integrated over the diffuse emission at frequencies 1.42 ($n=166$), 1.28 ($n=172$) 
and 0.61 ($n=220$) GHz.  
The feature detected at $\sim 100$ kms$^{-1}$ is the hydrogen line. 
Gaussians were fitted to the hydrogen features at the observed frequencies and
the results are shown in Table \ref{tab4}.  
We did not detect carbon RRL from G24.83+0.10 at any of the observed frequencies.
If the carbon lines were formed in an associated PDR with $T_e \le 400$ K,
$n_e > 3$ cm$^{-3}$ and pathlengths $ \ge 11$ pc, we should have been able
to detect these near 1 GHz, with the sensitivity
of our present observations. 

\begin{table*}
\caption{Gaussian fits to the source-integrated line emission from the sources
in the G24 region.  The $1\sigma$ errors on the fitted parameters are given. } 
\begin{tabular}{ccllll}
\hline
& &   & & &  \\
Source & Line &   S$_{int}$  &  $V_{LSR}$  & FWHM    & Reference   \\
          &          &  mJy         &   kms$^{-1}$     & kms$^{-1}$     &   \\
\hline
G24.83+0.10 &  H$166\alpha$  & $153\pm15$  & $109\pm4$  & $20\pm8$  & A \\
 &  H$172\alpha$   & $179\pm13$  & $104.5\pm1$  & $30\pm1.5$   & A   \\  
 &  H$220\alpha$   &$250\pm39$  & $107\pm1.5$ & $19\pm2$  &  A   \\
 &  H$87\alpha^1$ &$370$  & $108.6\pm0.4$ & $26.9\pm0.8$  & B  \\
 &  H$110\alpha^2$  & 164 & $107$ & 23 &  C  \\
 &  H$272\alpha^3$  & 235  & $83\pm2$ & $77\pm4$ & D  \\
 &  C$272\alpha^3$  &   & $46\pm2$ & $34\pm5$ & D  \\
\hline
\end{tabular}

$^1$$~~$ integrated over $3'$; 
$^2$$~~$ integrated over $2.6'$; \\
$^3$$~~$ integrated over $2^{\circ} \times 6'$. \\
A.$~~$ Present work; $~~~~$  
B.$~~$ Lockman (1989) \nocite{lockman} \\
C.$~~$ Downes et al. (1980) \nocite{downes}; $~~~~$
D.$~~$ Anantharamaiah (1985) \nocite{anantha} \\

\label{tab4}
\end{table*}

\begin{figure}
\includegraphics[width=4in]{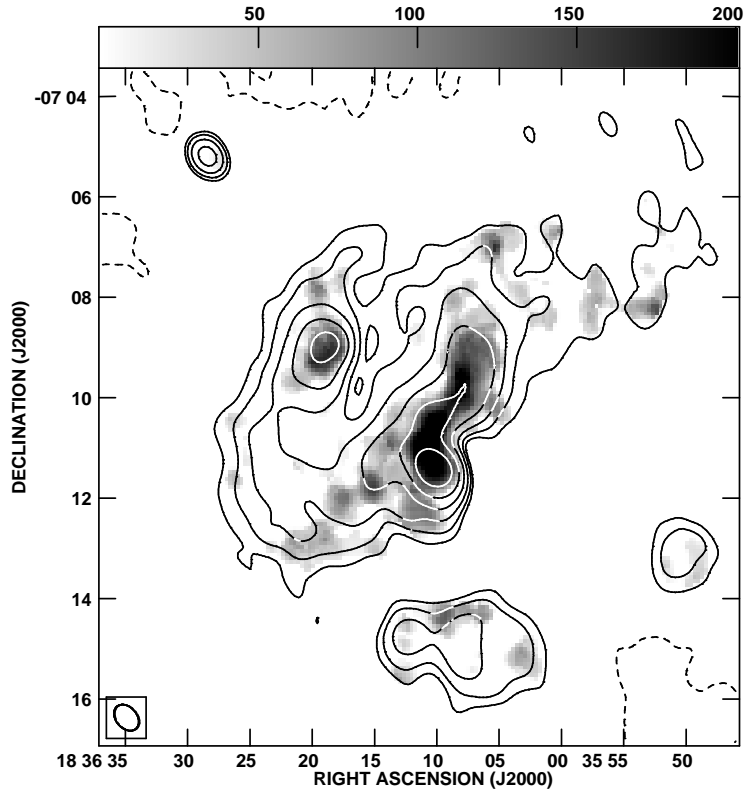}
\caption{ VLA Moment 0 image of H$166\alpha$ (grey scale) of the HII
region G24.83+0.10 superposed on a contour
map of the continuum emission from the region at 1.42 GHz.  The image has a resolution
of $35''\times 25''$ PA$=40^{\circ}$.  }
\label{fig9a}
\end{figure}

The integrated line intensity map of the H$166\alpha$ emission (in grey) is shown in 
Fig \ref{fig9a}  along with the radio continuum (in contours) at 1.42 GHz.
Line emission from G24.83+0.10 follows the continuum morphology.  Strongest line
emission is observed near the compact objects G24.80+0.10 and and G24.85+0.09. 
Line emission is also observed from the diffuse source (see Fig \ref{fig9a}).
The line emission arising in G24.83+0.10 is observed at velocities ranging from
$\sim 100$ to 110 kms$^{-1}$.  No systematic gradient in the velocity is observed.  
Higher velocities $\sim 115-125$ kms$^{-1}$ are observed 
for the RRL arising near the UCHII G24.80+0.09.  From their high resolution CO observations,
Furuya et al. (2002) \nocite{furuya}  find that
the LSR velocity of the molecular cloud is about 111 kms$^{-1}$ with velocity in
the outflow ranging from 90 to 130 kms$^{-1}$.  Thus, the
RRL we detect near this UCHII region is likely related to this star forming region. 
Interestingly, the lines appear to be narrow 
with widths at half maximum of $12-24$ kms$^{-1}$.
The narrow lines seem to arise close to the HII region G24.80+0.10 whereas the widest 
lines are observed near the UC HII region G24.85+0.09.  

Weak H RRL emission from
diffuse HII regions G24.74+0.08 and G24.74+0.16 
is also detected (see Fig \ref{fig9a}) at 102 kms$^{-1}$ and 106 kms$^{-1}$ respectively.

Due to the low signal-to-noise ratio for the H$220\alpha$ and H$172\alpha$ data,
we were unable to obtain a detailed distribution of emission across
G24.80+0.01. Thus, 
the integrated line flux densities of G24.83+0.10 
at the three observed frequencies were used to model the line emission.  
The departure coefficients required in the models were calculated using
the computer code of Salem \& Brocklehurst (1979) \nocite{salem}, modified by Walmsley
\& Watson (1982) \nocite{walmsley} and 
Payne, Anantharamaiah and Erickson (1994) \nocite{payne}.
These were then used to calculate the expected line flux density at the
different quantum numbers following Shaver (1975) \nocite{shaver2}.

The best fitting model that we found had the following parameters, typical
of a diffuse HII region:
$T_e=7000$ K and $n_e=100$ cm$^{-3}$ and a line of sight extent of 11 pc. 
The background radiation field for all the models was assumed to be
a HII region with an emission measure of $5\times10^4$ pc~cm$^{-6}$.
However we note that this model overpredicts the line flux near 0.61 GHz.

Detailed modelling of the observed RRLs should include
contribution from both the diffuse
HII region and the compact HII regions observed embedded within the diffuse shell.
With the signal to noise of the current data, such a detailed model is
not warranted.  High resolution sensitive low frequency
RRL observations are required to model different regions of G24.80+0.10 independently.

\section{Summary}
Using the GMRT and VLA, we have carried out low frequency interferometric 
continuum and recombination line observations of
the HII regions near $l=24^{\circ}.8$, $b=0^{\circ}.1$ at
0.61 GHz, 1.28 GHz and 1.42 GHz.  We detect continuum emission from
ultra-compact, compact and extended HII regions.  In this paper, we
discuss the morphology and average properties of the HII regions
using our continuum observations. 
G24.83+0.10, the main region under study in the paper, is diffuse and massive
encompassing $\sim 550~M_{\odot}$ within a linear size of 7 pc with a rms electron
density of $\sim 100$ cm$^{-3}$.  The source has
a shell-like morphology  with the shell opening to the N.  Two IRAS point sources
are clearly associated with this HII region. 
Compact HII regions are embedded in this diffuse region.  

We detected H$220\alpha$, H$172\alpha$ and H$166\alpha$ lines from G24.83+0.10. 
We also detected RRLs from two other HII regions in the field near 100 kms$^{-1}$. 
The model which provides the best fit to the source-integrated RRL data towards
G24.83+0.1 has the following parameters:
$T_e =7000 $ K, $n_e \sim 100$ cm$^{-3}$ and path length of 11 pc.
However this model overestimates the line strength at 0.61 GHz.
We note that the electron
density of the diffuse HII region that best fits the observed line data
is similar to the rms electron density found from the continuum observations.
More sensitive RRL
observations of G24.83+0.10 are required to develop a detailed model of emission
for this HII region.  

\paragraph {Acknowledgments:}
We thank the anonymous referee for several useful comments which have especially
helped improve the IR part of the
paper.  We thank the staff of the GMRT that made these observations possible.
GMRT is run by the National Centre for Radio Astrophysics of the Tata Institute
of Fundamental Research.  The National Radio Astronomy Observatory is a facility
of the National Science Foundation operated under cooperative agreement by Associated
Universities, Inc.  NGK thanks Prasad Subramanian for useful comments on the manuscript.
The IRAS point source catalog was obtained from VizieR.

\end{document}